\documentclass[aps,prl,twocolumn,superscriptaddress,10pt,floatfix]{revtex4-2}
\usepackage{symbols}
\usepackage{ORI_Group_Style}
\usepackage[english]{babel}

\usepackage{relsize}
\usepackage{xcolor}
\usepackage{scalerel}
\usepackage{tikz}
\usetikzlibrary{svg.path}
\usepackage{soul}
\definecolor{orcidlogocol}{HTML}{A6CE39}
\tikzset{
  orcidlogo/.pic={
    \fill[orcidlogocol] svg{M256,128c0,70.7-57.3,128-128,128C57.3,256,0,198.7,0,128C0,57.3,57.3,0,128,0C198.7,0,256,57.3,256,128z};
    \fill[white] svg{M86.3,186.2H70.9V79.1h15.4v48.4V186.2z}
                 svg{M108.9,79.1h41.6c39.6,0,57,28.3,57,53.6c0,27.5-21.5,53.6-56.8,53.6h-41.8V79.1z M124.3,172.4h24.5c34.9,0,42.9-26.5,42.9-39.7c0-21.5-13.7-39.7-43.7-39.7h-23.7V172.4z}
                 svg{M88.7,56.8c0,5.5-4.5,10.1-10.1,10.1c-5.6,0-10.1-4.6-10.1-10.1c0-5.6,4.5-10.1,10.1-10.1C84.2,46.7,88.7,51.3,88.7,56.8z};
  }
}
\newcommand\orcid[1]{\href{https://orcid.org/#1}{\mbox{\scalerel*{
\begin{tikzpicture}[yscale=-1,transform shape]
\pic{orcidlogo};
\end{tikzpicture}
\ 
}{|}}}}

\hypersetup{colorlinks,linkcolor=blue,citecolor=blue,urlcolor=blue}

\begin{document}

\title{Macroscopic Quantum Superpositions via Dynamics in a Wide Double-Well Potential}

\author{M. Roda-Llordes
}
\affiliation{Institute for Quantum Optics and Quantum Information of the Austrian Academy of Sciences, 6020 Innsbruck, Austria}
\affiliation{Institute for Theoretical Physics, University of Innsbruck, 6020 Innsbruck, Austria}
\author{A. Riera-Campeny
}
\affiliation{Institute for Quantum Optics and Quantum Information of the Austrian Academy of Sciences, 6020 Innsbruck, Austria}
\affiliation{Institute for Theoretical Physics, University of Innsbruck, 6020 Innsbruck, Austria}
\author{D. Candoli}
\affiliation{Institute for Quantum Optics and Quantum Information of the Austrian Academy of Sciences, 6020 Innsbruck, Austria}
\affiliation{Institute for Theoretical Physics, University of Innsbruck, 6020 Innsbruck, Austria}
\author{P. T. Grochowski
}
\affiliation{Institute for Quantum Optics and Quantum Information of the Austrian Academy of Sciences, 6020 Innsbruck, Austria}
\affiliation{Institute for Theoretical Physics, University of Innsbruck, 6020 Innsbruck, Austria}
\affiliation{Center for Theoretical Physics, Polish Academy of Sciences, Aleja Lotnik\'ow 32/46, 02-668 Warsaw, Poland}
\author{O. Romero-Isart
}
\affiliation{Institute for Quantum Optics and Quantum Information of the Austrian Academy of Sciences, 6020 Innsbruck, Austria}
\affiliation{Institute for Theoretical Physics, University of Innsbruck, 6020 Innsbruck, Austria}

\date{\today}

\begin{abstract}

We present an experimental proposal for the rapid preparation of the center of mass of a levitated particle in a macroscopic quantum state, that is a state  delocalized over a length scale much larger than its zero-point motion and that has no classical analog. This state is prepared by letting the particle evolve in a static double-well potential after a sudden switchoff of the harmonic trap, following initial center-of-mass cooling to a sufficiently pure quantum state. We provide a thorough analysis of the noise and decoherence that is relevant to current experiments with levitated nano- and microparticles.  In this context, we highlight the possibility of using two particles, one evolving in each potential well, to mitigate the impact of collective sources of noise and decoherence. The generality and scalability of our proposal make it suitable for implementation with a wide range of systems, including single atoms, ions, and Bose-Einstein condensates.  Our results have the potential to enable the generation of macroscopic quantum states at unprecedented scales of length and mass, thereby paving the way for experimental exploration of the gravitational field generated by a source mass in a delocalized quantum state.

\end{abstract}

\maketitle

Over the last century, many efforts have been directed towards preparing a delocalized state of increasingly massive objects over a distance comparable to their size \cite{Davisson1927,vonHalban1936,Keith1988,Arndt1999,Fein2019}. The preparation of such macroscopic quantum superposition states of massive particles holds significant interest across the field of quantum science~\cite{Hornberger2012}. Their high susceptibility to external stimuli equips them with excellent sensor capabilities. They also provide a testing ground for collapse models~\cite{Ghirardi1986, Ghirardi1990, RomeroIsart2011, Bassi2013}, which predict the breakdown of the quantum superposition principle at large scales.  Moreover, they could enable the direct observation of the gravitational field generated by a sufficiently large source mass in a quantum superposition state, which would shed light upon the interplay between quantum mechanics and gravity~\cite{Rickles2011,Belenchia2018}. 
The preparation of such macroscopic quantum states requires:
(i) Fast experimental runs to avoid collisions with gas molecules~\cite{Joos1985,RomeroIsart2011,Weiss2021,Neumeier2022}, (ii) Minimal use of laser light to avoid decoherence due to photon scattering~\cite{Jain2016,Pino2018,Maurer2023} and internal particle heating, which critically determines decoherence due to thermal emission~\cite{Hackermuller2004,RomeroIsart2011}, (iii) Access to nonlinearities to generate negative Wigner function states, and (iv) The ability to repeat nearly identical experimental runs quickly and with the same particle to avoid low-frequency noise, drifts, and other systematic errors.

\begin{figure}[t]
    \centering
    \includegraphics[width=\linewidth]{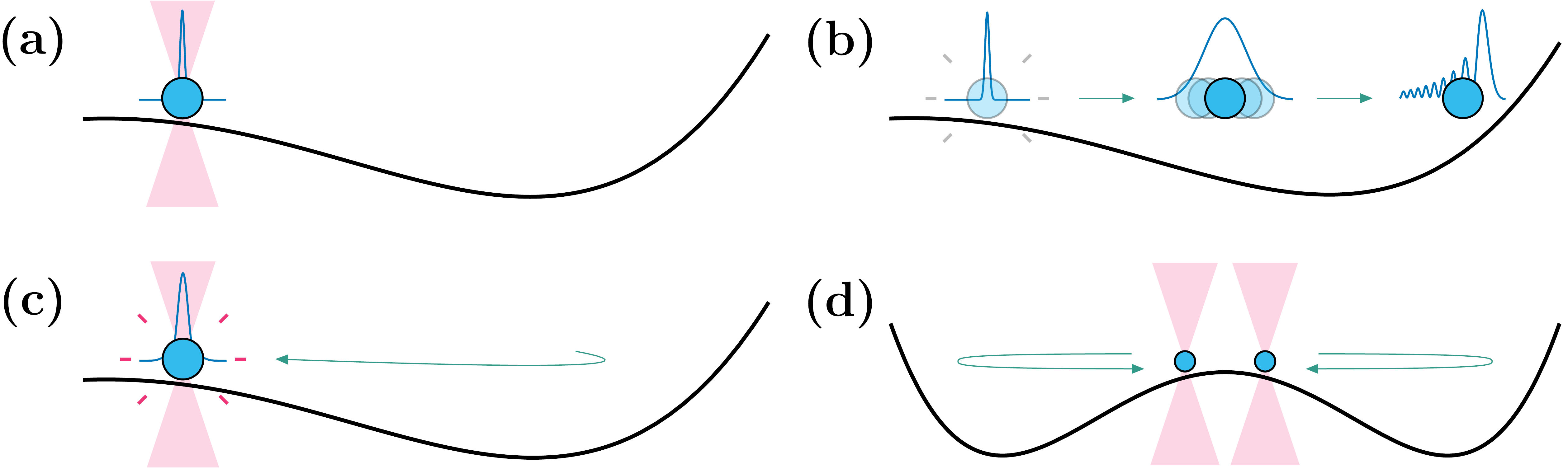} 
    \caption{
    Schematic representation of the protocol.
    (a)~The particle is initially trapped and cooled in a harmonic potential. 
    (b)~The trap is switched off and the particle explores the nonharmonic potential, experiencing both coherent expansion and non-Gaussian physics.
    (c)~The particle returns to the original position, allowing for repetition of the protocol.
    (d)~Extending the protocol to two particles allows for collective noise mitigation and detection of weak interactions.
    }
    \label{figure1}
\end{figure}

In this paper, we propose a scheme for preparing macroscopic quantum superposition states that simultaneously satisfies the challenging requirements (i)-(iv). This scheme is based on levitation and control of micro-objects in high vacuum~\cite{Gonzalez-Ballestero2021}. The scheme exploits the quantum nonlinear dynamics generated in a static nonharmonic potential (e.g.,~double-well potential, see~\figref{figure1}), which is assumed to be wide, time-independent, and implemented with static nonoptical fields. The dynamics is triggered after switching off a tighter harmonic potential (e.g.,~optical trap) where center-of-mass cooling is performed~\cite{Delic2020,Magrini2021,Tebbenjohanns2021,Kamba2022,Ranfagni2022, Piotrowski2023,Kamba2023}. The harmonic potential is centered near the top of the double-well potential but sufficiently far (compared to the wave-function size) so that the induced dynamics occurs in one of the wells only [\figref{figure1}(a)]. Because of the wide-ranging size of the double-well potential particle quantum tunneling is absent. This nonharmonic potential is convenient as it induces both coherent inflation~\cite{Romero-Isart2017,Pino2018}, namely an exponentially fast generation of motional squeezing via the inverted harmonic term, and non-Gaussian physics when the particle wave packet arrives at the turning point where the quartic term of the potential dominates [\figref{figure1}(b)]. Note that the particle will evolve very rapidly and perform a loop, returning to the initial position where the harmonic potential can be switched on again to repeat the experimental run [\figref{figure1}(c)]. The double-well potential is also convenient as our protocol can be extended to two particles, one evolving in each of the wells in a mirror-symmetric way [\figref{figure1}(d)]. This is useful as the two-particle dynamics can be used to mitigate collective sources of noise and decoherence by performing differential measurements~\cite{Pedernales2022} as well as to detect weak long-range interactions between them~\cite{Weiss2021,Cosco2021,Rieser2022}.

More specifically, we consider the center-of-mass motion of a particle of mass $\mass$ and focus on the motion along a given axis, described by the particle's center-of-mass position and momentum operators $\xOu$ and $\pOu$ fulfilling $\coms{\xOu}{\pOu}= \im \hbar$. The possible cross-coupling to other center-of-mass degrees of freedom is assumed to add noise and decoherence, whose effect is analyzed later. For times $t< 0$ we assume that the particle is cooled to a thermal state of a harmonic potential of frequency $\trapfreq$ centered at position $\xstartu$, namely $\Vtrap (\xu)= \mass \trapfreq^2 (\xu-\xstartu)^2/2$. A thermal state is characterized by its phonon mean number occupation $\bar n$. Today, it is experimentally feasible to cool a levitated dielectric nanoparticle to the ground state ($\bar n<1$)~\cite{Delic2020,Magrini2021,Tebbenjohanns2021,Kamba2022,Ranfagni2022, Piotrowski2023,Kamba2023} with  position and momentum zero-point fluctuations given by $\xzpf=\spare{\hbar/(2\mass\trapfreq)}^{1/2}$ and $\pzpf=\hbar/(2\xzpf)$, respectively. At $t=0$, the harmonic potential is switched off (e.g., optical trap is turned off) such that a weaker nonharmonic potential in the background (e.g., generated by electrostatic fields~\cite{Millen2015,Conangla2020,Dania2021}) is dominant. The center-of-mass quantum dynamics generated by this nonharmonic background potential is the focus of this paper. We consider the double-well potential $\Vdw(\xu)= \mass\dwfreq^2 [-\xu^2 + \xu^4/(2\dwdistanceu^2)]/2$, parameterized by the frequency $\dwfreq$ and length $\dwdistanceu$. We remark that in absence of noise and decoherence, the induced dynamics and the corresponding generated quantum states depend on the following six parameters: $\mass$, $\trapfreq$, $\xstartu$, $\bar n$, $\dwfreq$, and $\dwdistanceu$.

In this paper, we will focus on the quantum dynamics generated in {\em wide} double-well potentials, i.e., potentials for which $\dwdistanceu \gg \xzpf$ and $\dwfreq \ll \trapfreq$.
The reason for this is that nano- and microparticles cooled to the ground state have subatomic zero-point motion fluctuations $\xzpf \ll 10^{-10} \text{ m}$ and double-well potentials generated via static fields have at least micrometer-sized scales $\dwdistanceu \gg 10^{-7} \text{ m}$~(for example, see~\cite{Ciampini2021} for an experimental implementation). 
In such wide double-well potentials we will focus on the quantum dynamics generated when the particle evolves in one of the two wells (say, the right one for $\xu>0$, see \figref{figure1}), which requires $\xstartu \gg \xzpf$.  In this parameter regime, large phase-space expansions associated with the generation of large motional squeezing are expected. The numerical simulation of the Wigner function in dynamical scenarios (including noise and decoherence) where large phase-space squeezing and non-Gaussian states are generated is challenging. 
In order to overcome this challenge, we have developed numerical and analytical methods that we present in~\cite{Roda-Llordes2023} and~\cite{Riera-Campeny2023}, respectively. These references, and specially \cite{Riera-Campeny2023}, provide the supplemental material and further analysis required to validate the results presented in this experimental proposal.

\begin{figure}[t]
    \centering
    \includegraphics[width=\linewidth]{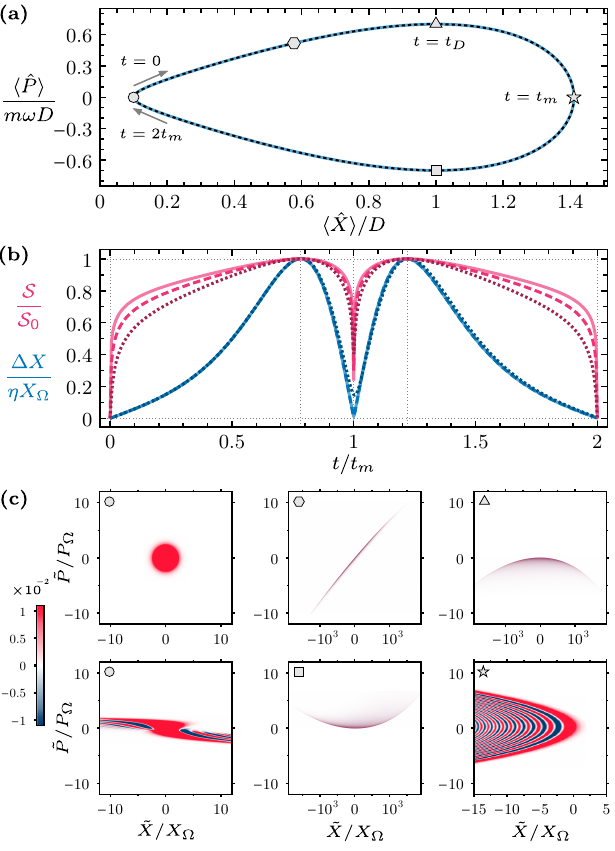}
    \caption{
    Coherent quantum dynamics.
    (a)~First order moments (solid line) and classical trajectory (dashed line).
    (b)~Normalized Gaussian motional squeezing and position variance for different double-well potentials.
    Solid, dashed, and dotted lines correspond to XXL, XL, and L in Table~\ref{table:parameters}.
    (c)~Wigner function for the L set of parameters at different moments of time, ordered clockwise and corresponding to the indicated points in panel (a).
    Numerical results obtained using the split-step method~\cite{Leforestier1991}.}
    \label{figure2}
\end{figure}

\begin{table}[b]
\caption{Double-well parameters considered in this paper. Configurations that generate quantum states with squeezing (in variance) of the order of $\{1,20,40,60,80,100\}$ decibels are defined as {\em sizes} \{XS, S, M, L, XL, XXL\}, respectively. \label{table:parameters}}
    \begin{tabular}{|c|c|c|c|c|c|}
    \hline
        Size & $\xstartu/\dwdistanceu$ & $\dwfreq/\trapfreq$ & $\dwdistanceu/\xzpf$ & $\sqrt{2}\,\eta$ &  $ \squeez_0 [\text{dB}] $  \\ \hline
        XXL & $10^{-1}$ & $10^{-4}$ & $10^8$ & $10^5$ & $97$  \\ 
        XL & $10^{-1}$ & $10^{-3}$ & $10^6$  & $10^4$ & $77$  \\ 
        L & $10^{-1}$ & $10^{-2}$ & $10^4$  & $10^3$ & $57$ \\  \hline
    \end{tabular} 
\end{table}

Let us now show the coherent dynamics generated in these wide double-well potentials.
To that end, we numerically solve the Wigner function dynamics for the sets of parameters given in Table~\ref{table:parameters} assuming $\bar n=0$ for now. Let us first focus on the quantum dynamics of the first and second phase-space moments. 
\figref{figure2}(a) shows the phase-space trajectory given by the dimensionless position and momentum expected values given by $\avg{\xOu}(t)/\dwdistanceu$ and $\avg{\pOu}(t)/(\mass \dwfreq \dwdistanceu)$, respectively. In these units, the trajectory is approximately the same for any of the set of parameters listed in~Table~\ref{table:parameters}. The trajectory followed by the expected values is nearly indistinguishable from the phase-space classical trajectory $\xuc(t)$ and $\puc(t)= \mass \dot \xuc(t)$ followed by a particle with the initial condition given by $\xuc(0)=\xstartu$ and $\puc(0) =0$, which has an analytical solution~\cite{Hsu1960, Brizard2009}. The trajectory is a closed orbit that facilitates the repetition of experimental runs. The orbiting period $2\tmax$ can be well approximated by $ \dwfreq\tmax =  \log(4\sqrt{2}\dwdistanceu /\xstartu)$~\cite{Hsu1960, Brizard2009}. In order to prevent decoherence due to the scattering of gas molecules we require that the probability to scatter a single gas molecule during an experimental run, that is during the orbiting time $2 \tmax$, is negligible. This condition is given by  $\tmax \ll \tgas/2$, where $\tgas$ is the timescale associated with a single gas scattering event and for a spherical particle of radius $R$ is given by $\tgas = 3\sqrt{\massgas \boltzmannconstant \temperaturegas}/(16\pi\sqrt{2\pi}\pressuregas \radius^2)$ ~\cite{RomeroIsart2011}, where $\massgas$, $\temperaturegas$, and $\pressuregas$ are the single molecule mass, temperature, and pressure of the gas, respectively. This important requirement can be satisfied in ultra-high vacuum, where $\tgas$ for nanoparticles is of the order of milliseconds.

Figure \ref{figure2}(b) shows the time dependence of the position standard deviation $\Delta \xu (t)=\spares{\avg{\xOu^2}(t)-\avg{\xOu}(t)^2}^{1/2}$ and the Gaussian motional squeezing $\squeez(t)$~\footnote{The motional Gaussian squeezing is given by $\squeez(t) = -10 \log_{10}[ \min_\theta \Delta^2 \xu_\theta(t)/\xzpf^2]$, where $\Delta \xu_\theta (t) = \spares{\avg{\xOu_\theta^2}(t)-\avg{\xOu_\theta}(t)^2}^{1/2}$ is the standard deviation of the rotated phase-space quadrature in length units, i.e., $\xOu_\theta = \cos (\theta) \xOu + \xzpf \sin (\theta) \pOu / \pzpf $. Importantly, $\Delta \xu_\theta (t) $ is evaluated using the quadratic dynamics only, see~\cite{Riera-Campeny2023} for further details.}. The quantum dynamics generates a large spatial quantum delocalization and motional squeezing. As one can observe in \figref{figure2}(b), maximum spatial delocalization is achieved at $t=\td$, when  $\avg{\xOu}(\td) = D$ and $\Delta \xu (\td)/\xzpf = \eta$ with 
\be
\eta = \frac{1}{\sqrt{2}}\frac{\trapfreq }{ \dwfreq  } \frac{ \dwdistanceu}{  \xstartu }.
\ee
The corresponding motional squeezing is given by $\squeez (\td) = \squeez_0$ where $\squeez_0 = -10 \log_{10}(\eta^{-2})$.
The parameter $\eta$ is thus key to quantifying the amount of spatial quantum delocalization and motional squeezing generated during the coherent dynamics in the double-well potential. Table~\ref{table:parameters} shows how spatial delocalization orders of magnitude larger than the zero-point motion with associated motional squeezing of several tens of decibels can be rapidly generated during the evolution in the double-well potential. This fast generation of motional squeezing is due to the coherent inflation generated by the inverted harmonic term present in the potential~\cite{Romero-Isart2017,Pino2018}.
At the turning points, $t=\tmax$ and $t=2 \tmax$, the quantum state recompresses such that the position and momentum fluctuations are of the order of the zero-point motion. The recompression is more effective the wider the size of the double-well potential, namely the larger the value of $\eta$. These expansion and compression dynamics resembles the loop protocol~\cite{Weiss2021} and facilitates the repetition of an experimental run. In contrast to~\cite{Weiss2021}, the generated motional squeezing enhances the effect of the nonlinearities in the double-well potential, thereby preparing quantum non-Gaussian states with Wigner negativities, as we explicitly show in the following.

\begin{figure}
    \centering
    \includegraphics[width=\linewidth]{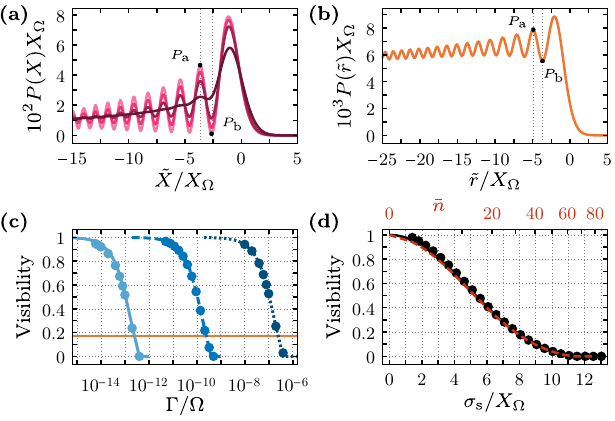}
    \caption{
    Quantum dynamics in the presence of decoherence.
    (a)~Position probability distribution at $t=\tmax$ $P(\xu,\tmax)$ for the L set of parameters (Table~\ref{table:parameters}) and with decoherence rates $\dnoiserate/\trapfreq=\{1,4,10,40\}\times 10^{-8}$. Darker lines correspond to higher $\dnoiserate/\trapfreq$.
    (b)~Relative position probability distribution~\cite{sm} at $t=\tmax$ $P(\tilde{r},\tmax)$ for the same potential.
    Visibility of the first minimum of $P(\xu,\tmax)$ as a function: (c)~$\dnoiserate/\trapfreq$ for XXL (solid), XL (dashed), and L (dotted) set of parameters. 
    The horizontal line corresponds to the visibility of the first minimum of $P(\tilde{r},\tmax)$ as a function of collective $\dnoiserate/\trapfreq$.
    (d)~Initial position imprecision $\sigma_\text{s}$ (solid line) and $\bar{n}$ (dashed line).
    Lines in (c), (b) and (d) correspond to results obtained using the analytical treatment presented in~\cite{Riera-Campeny2023}, while all other results are obtained from numerical simulation using the method presented in~\cite{Roda-Llordes2023}.}
    \label{figure3}
\end{figure}

In \figref{figure2}(c) we plot the Wigner function at the six relevant instances of time indicated in \figref{figure2}(a), using the L set of parameters (see Table~\ref{table:parameters}). The Wigner function is represented with phase-space coordinates $\tilde \xu$ and $\tilde \pu $ centered at the classical trajectory, namely $\tilde \xu = \xu - \xuc(t)$ and $ \tilde \pu = \pu - \puc(t)$. One can observe that when the quantum state is  largely squeezed, the potential does not only rotate the squeezed state in phase space, as the harmonic part of the potential does, but the nonharmonic terms bend the phase-space distribution in a boomeranglike shape, thereby generating Wigner negativities and interference fringes~\cite{Zurek2001,Stobinska2008}. These boomeranglike states can be well described by a cubic-phase state~\cite{Weedbrook2012,Brunelli2019,Moore2022,Neumeier2022,Kala2022}, that is, the state obtained by applying a cubic-phase operator to a squeezed state. At the turning point $t=\tmax$, the cubic-phase state is such that  an interference pattern in the probability position distribution is obtained, see \figref{figure3}(a). Using the analytical tools derived in~\cite{Riera-Campeny2023}, one can show that this probability distribution is given by a squared Airy function with a fringe separation between the two first maxima $\fringesep$ that scales as $\fringesep /\xzpf \sim (\trapfreq/\dwfreq)^{2/3} (\xzpf/\dwdistanceu)^{1/3}$, for a fixed $\xstartu/\dwdistanceu$. For the set of parameters given in Table~\ref{table:parameters}, $\fringesep /\xzpf \approx 2.5$ for all cases. Finally, one can also show~\cite{Riera-Campeny2023} that the state after one period, that is, at time $t=2\tmax$, can be approximated as a quartic-phase state, namely a state obtained by applying a quartic-phase operator to a coherent state.

Before discussing how the generation of these macroscopic quantum states can be certified, let us discuss the impact of noise and decoherence. We emphasize that during the dynamics no laser light is used, and hence decoherence due to recoil heating is absent~\cite{Jain2016,Maurer2023}. In addition, under the regime, $2 \tmax \ll \tgas$, achieved in ultra-high vacuum~\cite{Dania2023} and fast experimental runs, decoherence due to the scattering of gas molecules is prevented. Hence, the main sources of noise and decoherence will be: (i)~Thermal emission from the particle~\cite{Hackermuller2004,RomeroIsart2011,Bateman2014, Agrenius2023}, (ii)~Fluctuations in the double-well potential, both in amplitude and position~\cite{Gehm1998,Henkel1999,Schneider1999}, (iii)~Fluctuating forces acting on the particle (e.g., due to fluctuating electric fields~\cite{Brownnutt2015,Kumph2016}), (iv)~Finite phonon number occupation ($\bar n>0$) and/or initial position imprecision (i.e., different $\xstartu$ in each experimental run), and (v)~Timing imprecision, i.e., different values of the measurement time $\tmax$ (or $2 \tmax$) in each experimental run. Cases (i)-(iii) can be modeled by calculating the dynamics using the master equation
\be
    \partial_t \rhoop(t) = -\frac{\im}{\hbar} \coms{\Hop}{\rhoop(t)} 
   - \frac{\dnoiserate}{2 \xzpf^2} \coms{\xOu}{\coms{\xOu}{\rhoop(t)}},
\ee
where $\dnoiserate = \dnoiserate_{T}+\dnoiserate_{P} + \dnoiserate_{F} $ is the decoherence with contributions from (i), (ii), and (iii). The expression of $\dnoiserate_{T}$ can be found in the literature~\cite{Schlosshauer2007,RomeroIsart2011,Bateman2014} and for the particular case of a silica nanoparticle with internal temperature $T$ and trap frequency $\trapfreq/(2 \pi)=100~\text{kHz}$ is given by $\dnoiserate_T/\trapfreq \approx 10^{-10} \times \spare{T /(300~\text{K})}^6$~\footnote{This expression assumes a frequency-independent permittivity given by $\epsilon=2.1 + 0.57 \im$}.
The expression of $\dnoiserate_{P}$ can be obtained by considering the potential fluctuations $\spares{1+\stochvartwo(t)}\Vdw(\xOu + \stochvarone(t) \xzpf)$, where $\stochvarj(t)$ for $j=1,2$ are dimensionless stochastic Gaussian variables of zero mean and assumed delta-correlated in the relevant timescales, namely $\avg{\stochvarj(t) \stochvarj(t')} = 2 \pi \psdj \delta(t-t')$. For weak fluctuations and when the particle is at $\xuc$, it experiences a fluctuating force given by $\stochvarone(t)\xzpf \Vdw''(\xuc(t))  +  \stochvartwo(t) \Vdw'(\xuc(t)) $. Since during the closed trajectory one has that
$\Vdw''(\xuc(t)) < 5 \mass \dwfreq^2$ and $\Vdw'(\xuc(t)) < \sqrt{2} \dwdistanceu \mass \dwfreq^2$, the decoherence rate after ensemble average \cite{vanKampen1974}, see further details in \cite{Riera-Campeny2023}, is upper bounded by
\be
\frac{\dnoiserate_P}{\trapfreq} \leq \frac{2\pi\dwfreq}{4} \left(\frac{\dwfreq}{\trapfreq}\right)^3\left[25 \psdone + 2\pare{ \frac{\dwdistanceu}{\xzpf}}^2\psdtwo \right].
\ee
Finally, $\dnoiserate_{F}$ is obtained by considering a fluctuating force $F(t)$ (e.g., fluctuating electrostatic force) of zero mean, assumed white in the relevant frequency range with correlations given by $\avg{\stochF(t) \stochF(t')} = 2 \pi \psdF \delta(t-t')$. The associated decoherence rate is given by $ \dnoiserate_{F} = 2 \pi \xzpf^2 \psdF /\hbar^2$.
In \figref{figure3}(c) we show how the visibility of the interference pattern generated at $t = \tmax$ [see \figref{figure3}(a)] depends on $\dnoiserate$ for double well sizes defined in Table~\ref{table:parameters}. As we show in~\cite{Riera-Campeny2023}, the effect of decoherence scales with $\dnoiserate \eta^2 /\dwfreq$, and hence the wider the double-well potential, the more motional squeezing is generated, and the more stringent the requirements in $\dnoiserate$. From \figref{figure3}(c) and Table~\ref{table:parameters} one can rapidly calculate the values of the $\psdone$, $\psdtwo$, and $\psdF$ that are needed in an experiment to generate a visible interference pattern. 
Regarding cases (iv) and (v), we define   $\placingError$ and $\timingError$ as the standard deviations of normally distributed random variables that model the error in the initial position of the particle and in the time of the measurement, respectively. In \figref{figure3} we plot the visibility of the interference pattern as a function of $\bar n$ and $\placingError$.  Note that one can tolerate an initial position imprecision of up to $\placingError \sim 10\xzpf$ or, equivalently, up to $\bar n \sim 40$ in the initial state. Ground-state cooling is thus not a strict requirement. We have analyzed that timing errors up to $\timingError \sim 10^{-2}\dwfreq^{-1}$, which are experimentally feasible, provide a visible interference pattern.

In order to certify the generation of the macroscopic quantum states during the dynamics in the double-well potential, several strategies can be used. The most unambiguous strategy is to measure the position interference pattern generated at $t=\tmax$ [see \figref{figure3}(a)] using an inverted optical potential with harmonic frequency $\trapfreq_\text{i}$. It is known~\cite{Romero-Isart2017,Pino2018,Neumeier2022} that this technique magnifies the interference pattern without compromising its visibility if the condition $\fringesep/\xzpf \gg  (\dnoiserate/\trapfreq_\text{i})^{1/2}$ is met, where $\dnoiserate$ is dominated by optical back-action noise (i.e., recoil heating)~\cite{Jain2016,Maurer2023}. Alternatively, one could consider performing quantum tomography of the state at $2 \tmax$ and show the preparation of a state with a negative Wigner function. 
Finally, measuring $\Delta \xu(t)$, which is very sensitive to external noise and decoherence, and comparing the result to the predicted coherent value [shown in \figref{figure2}(b)] could be used as a method to certify that the overall dynamics was coherent.

As mentioned in the introduction and further analyzed in~\cite{sm}, one can consider the use of two particles, one evolving in each well in a mirror-symmetric way as illustrated in \figref{figure1}(d). We define the relative distance between the particles as $\tilde{r}=\tilde{\xu}_1-\tilde{\xu}_2$, where $\tilde{\xu}_i$ is the position of each particle relative to its classical trajectory. As shown in \figref{figure3}(b), its probability distribution at $t=\tmax$ also shows an interference pattern. While the visibility of this interference pattern is not one, even in the absence of noise and decoherence, it will be robust in front of sources of noise and decoherence that are collective, that is, that only affect the center-of-mass motion of the two particles~\cite{Pedernales2022}. Examples of this collective noise are fluctuations in the center of the double-well potential (e.g., due to vibrations) as well as the imprecision of the position of the two harmonic traps, whose separation is assumed fixed, with respect to the double-well potential. The latter could be implemented by using a standing-wave optical trap~\cite{Kamba2021}, such that the distance between two trapping points is fixed by the laser wavelength, or by using a programmable array of optical tweezers  with two nanoparticles~\cite{Vijayan2022,Vijayan2023,Rieser2022}. The standing-wave optical configuration can also be used to make sure that at $t=\tmax$ the particle is placed at a point of the standing wave where it experiences an inverted potential, which, as described above, is required to measure the interference pattern. Finally, the joint quantum dynamics of the two particles will be very sensitive to any weak interaction between them, and hence it could be used to detect weak interacting forces similarly to what is discussed in~\cite{Weiss2021,Cosco2021}. 

To conclude, we have shown how the dynamics of a massive particle in a wide nonharmonic potential can be used to rapidly prepare largely delocalized quantum states and a quantum interference pattern. In essence, our protocol implements an in-trap single particle matter-wave interference experiment ({\em \`a la} double-slit Young's experiment) in a way that circumvents key challenges for large masses, such as repeatability and absence of decoherence due to scattering of gas molecules. While observing macroscopic quantum physics is challenging~\cite{RomeroIsart2011,Romero-Isart2011,Yin2013,Scala2013,Bateman2014,Wan2016,Romero-Isart2017,Pino2018,Stickler2018,Weiss2021,Neumeier2022}, our results show what is required in terms of noise and decoherence and provide a  feasible path to scale up the mass of the objects that could be prepared in a macroscopic quantum superposition state. Our proposal is compatible with current state-of-the-art technology, such as optically trapped dielectric nanoparticles hybridized with electrostatic potentials~\cite{Millen2015,Conangla2020,Dania2021} or magnetically levitated superconducting spheres~\cite{Hofer2022,Latorre2022}. Finally, we emphasize that our scheme is scale-free and versatile and could thus be initially tested with  single atoms~\cite{Folling2007,Kaufman2015,Murmann2015,Dai2016,Brown2023}, ions~\cite{Harlander2011,Brown2011}, Bose-Einstein condensates~\cite{Shin2004,Albiez2005,Lesanovsky2006,Schumm2006,Bonneau2018,Borselli2021}, clamped nanomechanical oscillators~\cite{Pistolesi2021}, or even with a superconducting quantum circuit~\cite{venkatraman2023}.

\begin{acknowledgments}
MRL and ARC contributed equally to this work.
We thank the Q-Xtreme synergy group for fruitful discussions. This research has been supported by the European Research Council (ERC) under the grant agreement No. [951234] (Q-Xtreme ERC-2020-SyG) and by the European Union’s Horizon 2020 research and innovation programme under grant agreement No. [863132] (IQLev).
PTG was partially supported by the Foundation for Polish Science (FNP).
\end{acknowledgments}

\bibliography{bibliography}

\end{document}